\documentclass[conference]{IEEEtran}
\usepackage{amsmath,amssymb,amsthm,mathrsfs,amsfonts,dsfont,stmaryrd,wasysym,marvosym}
\usepackage{mathtools}
\usepackage{algorithm}
\usepackage{algorithmic}
\usepackage{color}
\usepackage{graphicx}  
\usepackage{psfrag, xcolor}
\usepackage{enumitem}
\usepackage{subfigure,cite,epsfig,url}
\usepackage{booktabs}
\allowdisplaybreaks[4]
\newtheorem{theorem}{Theorem}
  
\newtheorem{corollary}{Corollary}  
\newtheorem{proposition}{Proposition}   
\newtheorem{definition}{Definition} 
\newtheorem{remark}{Remark} 
\newtheorem{example}{Example} 
\newcommand{\mc}{\mathcal} 

\graphicspath{{figs/}}
\newcommand{\M}{\mathsf{M}}
\newcommand{\W}{\mathsf{W}}
\renewcommand{\L}{\mathsf{L}}
\newcommand{\K}{\mathsf{K}}
\newcommand{\tp}[1]{\textnormal{tp}(#1)}

\begin{document}
\fontencoding{OT1}\fontsize{10}{11}\selectfont

\title{Distributed Hypothesis Testing with Concurrent Detections}

\author{Pierre Escamilla$^{\dagger}$$^{\star}$\qquad Mich\`ele Wigger $^{\star}$ \qquad Abdellatif Zaidi$^{\dagger}$ $^{\ddagger}$  \vspace{0.3cm}\\
$^{\dagger}$ Paris Research Center, Huawei Technologies, Boulogne-Billancourt, 92100, France\\
$^{\ddagger}$ Universit\'e Paris-Est, Champs-sur-Marne, 77454, France\\
$^{\star}$ LTCI, T\'el\'ecom ParisTech, Universit\'e Paris-Saclay, 75013 Paris, France\\
\{\tt pierre.escamilla@huawei.com, abdellatif.zaidi@u-pem.fr\}\\
\{\tt michele.wigger@telecom-paristech.fr\}
}

\maketitle

\title{Coding For Multiple Possible Testing}

\begin{abstract}
A detection system with a single sensor and $\K$ detectors is considered, where each of the terminals observes a memoryless source sequence and the sensor sends a common message to all the detectors. The communication of this message is assumed error-free but rate-limited. The joint probability mass function (pmf) of the  source sequences observed at the terminals depends on an $\M$-ary hypothesis $(\M \geq \K)$, and the goal of the communication is that each  detector can guess the underlying hypothesis. Each detector $k$ aims to maximize the error exponent \emph{under hypothesis $k$}, while ensuring a small probability of error  under all other hypotheses. This paper presents an achievable exponents region for the case of positive communication rate, and characterizes the optimal exponents region for the case of zero communication rate. All results extend also to a composite hypothesis testing scenario.

\end{abstract}

\section{Introduction}~\label{secI}
\begin{figure}[!t]
	\begin{center}
		\includegraphics[width=.859\linewidth]{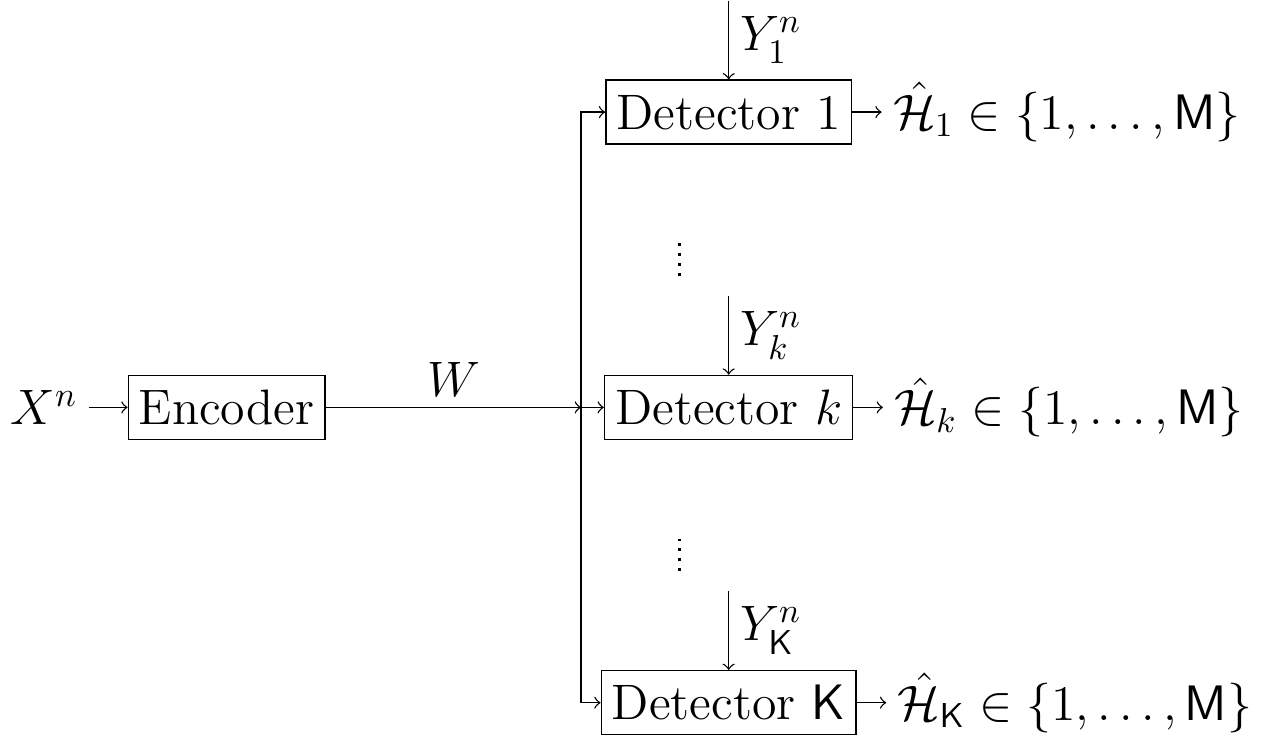}
		\caption{Distributed hypothesis testing with multiple detectors.}
		\label{fig-system-model}
	\end{center}
	\vspace{-5mm}
\end{figure}

Consider the multiterminal hypothesis testing scenario in Figure~\ref{fig-system-model} where an encoder observes a discrete memoryless source sequence $X^n\triangleq (X_1,\ldots, X_n)$  and communicates with multiple detectors over a \emph{common} noise-free bit-pipe of rate $R\geq 0$. Here, $n$ is a positive integer that  denotes the blocklength. Each detector $k \in \{ 1, \ldots \K\}$ observes a  memoryless source sequence $Y^n_k\triangleq(Y_{k,1},\ldots ,Y_{k,n})$, where the sequence of observations $\{(X_{t}, Y_{1,t}, Y_{2,t}, \ldots, Y_{\K,t} )\}_{t=1}^n$ is independent and identically distributed (i.i.d) according to a joint probability mass function (pmf) that is determined by the  hypothesis $\mc H \in \{1, \ldots, \mathsf{M}\}$. Under   hypothesis $\mc H=m$: 
\begin{equation}\label{eq:H}
 \{(X_{t}, Y_{1,t}, Y_{2,t}, \ldots Y_{\K,t})\}_{t=1}^n \textnormal{  i.i.d. } \sim P^{(m)}_{XY_1Y_2\ldots Y_\K},
 \end{equation}
We assume  $\M \geq \K$. Each detector~$k \in\{1,2,\ldots,\K\}$ decides on a  hypothesis,  $\hat{\mc H} \in\{1,2,\ldots,\M\}$,  with the goal to  maximize the  exponential decrease  of the probability of type-II error (i.e., of guessing $\hat{\mc {H}} \neq k$   when $\mc H=k$), while ensuring that the probabilities of type-I error (i.e., guessing  $\hat{\mc {H}}\neq m$  when $\mc H=m$ for some $m\neq k$) do not exceed a constant value $\epsilon_k \in(0,1)$ for all sufficiently large blocklengths $n$.  

The described scenario models, e.g., a
biometric system where $\K$ distinct accounts require authentication from users using biometric features. Each account $k$ aims to grant access to \emph{a different user} while denying access to all other users and guaranteing a given probability $\epsilon_k$ of false accept.  

Problems of distributed hypothesis testing are strongly rooted in both statistics and information theory.
In particular, the problem described above with only a single detector and two hypotheses ($\K=1$ and $\mathsf{M}=2$) was studied in \cite{AC86, han_hypothesis_1987,shimokawa,wagner}. The optimal exponent was derived for testing against conditional independence \cite{wagner}, but it remains open in general. 
Extensions to multiple terminals, multi-hop or interactive communication were presented in \cite{han_hypothesis_1987,zhao2014distributed,zhao2015distributed, katz2016collaborative, xiang2012interactive}. Moreover, \cite{sadaf_BC} studies testing against conditional independence with two detectors that  aim at maximizing the error exponent under the same hypothesis. The results in \cite{sadaf_BC} report a tradeoff between the exponents achieved at the two detectors when the sensor sends a common message of positive rate $R>0$ to both sensors. 
As this paper shows, such a tradeoff does not necessarily arise when the two detectors aim at maximizing the error exponents under two different hypotheses. (See Remark~\ref{remark: equivalence to non-adversarial many detectors}.)

In this paper, we propose a coding and testing scheme for a general number of hypotheses $\M$ and detectors $\K$  and positive rate $R>0$. In our scheme,
the sensor makes a tentative guess on the hypothesis and then applies the coding and testing scheme  that is adapted to the guessed hypothesis. Based on this scheme, we describe a set of achievable error exponents for given rate $R>0$.

Furthermore, we   exactly characterize the set of achievable error exponents for  communication rate $R=0$. That means when the sensor can send only a sublinear number of bits to the $\K$ detectors.  If the sensor  sends at most $\log(\M)$ bits, a tradeoff arises between their error exponents. In contrast, if  it  sends more than $\log(\M)$ bits, each detector can achieve the same error exponent as if the other detectors were not present. 
	
At last, we extend our results to a \emph{composite hypothesis testing} scenario where each receiver merges a set of hypotheses and aims at maximizing the error exponent under this set of hypotheses. This scenario models situations where a detector  wishes to raise an alarm in multiple events, without the need to specify which of the events occurred. It can also model uncertainty about the distribution implied by the  hypotheses. Error exponents for zero-rate single-detector composite hypothesis testing problems have previously been presented in \cite{shalaby1992multiterminal}.  

\textit{Notation:} The set of pmfs over an alphabet $\mathcal{X}$ is denoted $\mathcal{P}(\mc X)$ and the $n$-fold product of a pmf $P_X$ by $P_X^{\otimes n}$. 
and the type of a sequence $x^n$ is denoted by $\text{tp}(x^n)$.
We write $\mathcal{T}_{\mu}(P_X)$ for the set of $\mu$-typical sequences with respect to $P_X$.
Both $D(P_X\|P_{\bar{X}})$ and $D(X\|\bar{X})$ denote the Kullback-Leiber divergence between two pmfs $P_X$ and $P_{\bar{X}}$ over the same alphabet $\mc X$. 
 For an integer number $I$, we  abbreviate $\{1,\ldots,I\}$ by $[1:I]$.
Finally, $H(\cdot)$ and $I(\cdot;\cdot)$ denote entropy and mutual information. Where $I_P(A;B)$ indicates that  $I(A;B)$ is computed according to $P$.

\section{ Formal Problem Statement }~\label{secII}

Let $(X^n, Y_1^n, \ldots, Y_K^n)$ be i.i.d. according to one of the pmfs $P_{XY_1\cdots Y_K}^{(1)}, \ldots, P_{XY_1\cdots Y_K}^{(\M)}$ as described in \eqref{eq:H}. 
For clarity of exposition, assume that for each $k\in\{1,\ldots,\K\}$, the pmfs $P_{XY_k}^{(1)}, \ldots, P_{XY_k}^{(\M)}$ are all different and   $P^{(m)}_{XY_k}  \ll P^{(k)}_{XY_k}$, i.e.,  $P^{(k)}_{XY_k}$ is positive for all elements where $P^{(m)}_{XY_k}(x,y_k)$ is positive. The encoder observes a source sequence $X^n$. It applies an encoding function \begin{equation}
	\label{eq:phi}
f_{n}\colon  \mc X^n \rightarrow \mc W \triangleq [1:\mathsf{W}_n]
\end{equation}
 to $X^n$:
\begin{equation}
W = f_{n}(X^n),
\end{equation}
 and sends the resulting message $W$ to all  detectors.
Detector $k$  obtains $W$ and  observes  $Y_k^n$. It applies a decision function: 
\begin{equation}\label{eq:psik}
g_{k,n} \colon \mc W \times \mc Y_k^n \rightarrow[1:\mathsf{M}]. 
\end{equation}
 to decide on the hypothesis 
\begin{equation}\label{eq:decide}
\hat{\mc H}_k \triangleq g_{k,n} (W, Y_k^n).
\end{equation}

Type-I and type-II error probabilities at detector $k$ are defined as:
\begin{align}
\alpha_{k,m,n} &\triangleq    \text{Pr}\big\{\hat{\mc H}_k \neq m \big|\mc H =m \} ,\quad m\in[1:\M]\backslash \{k\}, \\
\beta_{k,n} &\triangleq   \text{Pr}\big\{\hat{\mc H}_k \neq k \big|\mc H =k \}.
\end{align}


\begin{definition}\label{def:1}
Given rate $R \geq 0$ and vector $\boldsymbol{\epsilon}=( \epsilon_1, \epsilon_2,\ldots,\epsilon_{\K})$ in $(0,1)^\K$, an error exponent vector $\boldsymbol{ \theta }=(\theta_1,\ldots,\theta_\K)$ is said achievable, if for each blocklength $n$ there exist  functions $f_{n}$, $g_{1,n},g_{2,n},\ldots,g_{\K,n}$  as in \eqref{eq:phi} and \eqref{eq:psik}, so that the following limits hold for all $k\in [1:\K]$:

\begin{IEEEeqnarray}{rCl}
\varlimsup_{n\to \infty} \alpha_{k,m,n} &\leq &\epsilon_k, \quad m\in[1:\M]\backslash \{k\},
\label{eq-definition-constant-constraints-typeI-errors}\\
  \varliminf_{n \to \infty}-\frac{1}{n} \log \beta_{k,n} &\geq&  \theta_k, \label{eq:thetak}
\end{IEEEeqnarray}
and 
\begin{align}
\varlimsup_{n\to \infty} \frac{1}{n}\log \mathsf{W}_n &\leq R.
\label{eq-definition-achievable-rates}
\end{align}
\end{definition}
For rate $R=0$ we make a finer distinction. Constraint $\mathsf{W}_n \leq \mathsf{W}$, for some constant value $\mathsf{W}$, is referred to as $R=0_{\mathsf{W}}$.
The main interest  of this document is on the set of all achievable error exponent vectors.
\begin{definition}\label{def:region}
Given  $R>0$ or $R=0_{\mathsf{W}}$ and  vector $\boldsymbol{\epsilon} \in (0,1)^\K$, the  closure of the set of all achievable exponent vector $\boldsymbol{\theta}=( \theta_1, \theta_2,\ldots,\theta_\K)$ is called the \emph{error exponents region $\mathcal{E}(R, \boldsymbol{\epsilon})$.} 
\end{definition}

\begin{remark}\label{remark: equivalence to non-adversarial many detectors} The error exponents region only depends on the sets of joint marginals $\{P_{XY_1}^{(m)}\}_{m=1}^{\mathsf{M}}$, $\{P_{XY_2}^{(m)}\}_{m=1}^{\mathsf{M}}$, $\ldots,$ $\{P_{XY_{\K}}^{(m)}\}_{m=1}^{\mathsf{M}}$. In our setup, each detector $k$ aims at maximizing the error exponent under hypothesis $\mc H =k$. When $P_X^{(1)}=P_X^{(2)}=\ldots = P_X^{(\K)}$, by relabelling above joint marginals, the setup is equivalent to a setup where all detectors aim at maximizing the error exponent under  hypothesis $\mc H=1$. In the special case of testing against conditional independence for $\K=2$ detectors and $\mathsf{M}=2$ hypotheses, this latter setup has been solved in \cite{sadaf_BC}. This optimal exponent is also achieved by our first main result, Theorem~\ref{theorem-lower-bounds-power-exponents-general-hypotheses-positive-rates}.
\end{remark}

\section{Main Results}
The following Theorem~\ref{theorem-lower-bounds-power-exponents-general-hypotheses-positive-rates} characterizes an inner bound to $\mathcal{E}(R,\boldsymbol{\epsilon})$. As we will see, it does not depend on $\boldsymbol{\epsilon}$.

Let for each positive rate $R>0$, $\mathcal{U}(R)$ be the set of tuples  $\big(P_{U|X}^{(1)}, \ldots, P_{U|X}^{(\M)}\big)$ satisfying 
\begin{equation}\label{eq:rate}
I_{P^{(m)}}( U; X|Y_k) \leq R, \quad k\in[1:\K], \; m\in[1:\M]\backslash \{k\},
\end{equation}
and 
\begin{equation}\label{eq:same}
P_{U|X}^{(m)} = P_{U|X}^{(m')}\quad  \textnormal{ if } \quad P_{X}^{(m)} = P_{X}^{(m')}.
\end{equation}
\begin{theorem}[Achievability under Positive Rate]~\label{theorem-lower-bounds-power-exponents-general-hypotheses-positive-rates} 
Given  $R > 0$ and $\boldsymbol{\epsilon}$ in $(0,1)^\K$, region $\mathcal{E}(R, \boldsymbol{\epsilon})$ contains all nonnegative vectors $\boldsymbol{\theta}=( \theta_1, \theta_2,\ldots,\theta_\K)$  that  satisfy the following two conditions for a tuple  $\big(P_{U|X}^{(1)}, \ldots, P_{U|X}^{(\M)}\big)\in \mathcal{U}(R)$:
\begin{align}\label{eq: an achievable rate exponent region for Heegard-Berger}
& \theta_k \leq \min_{\substack{m \in[1:\M] \\ m\neq k} }\;\;
 \min_{\substack{ \pi_{UXY_k} \colon  \\  \pi_{UX}  = P^{(m)}_{UX}  \\ \pi_{UY_k} = P_{UY_k}^{(m)}  }} 
 D\left(\pi_{UXY_k}||P^{(m)}_{U|X}P_{XY_k}^{(k)}\right)
\end{align}
and 
\begin{align}
\theta_k &\leq \min_{\substack{m \in [1 : \M]\\ m\neq k} }\;\; \min_{\substack{ \pi_{UXY_k} \colon  \\  \pi_{UX}  = P^{(m)}_{UX}  \\ \pi_{Y_k} = P_{Y_k}^{(m)} \\H_{\! P^{(m)}\!}(U|Y_k)\leq H_{\pi}(U|Y_k)}} \nonumber\\[1.2ex] &\Big[ D\left(\pi_{UXY_k}||P_{U|X}^{(m)}P_{XY_k}^{(k)}\right)  +R- I_{P^{(m)}}( U; X|Y_k) \Big].
\end{align}
\end{theorem}
\begin{IEEEproof}
See Section~\ref{secV}. 
\end{IEEEproof}
\begin{corollary}Assume that $\K=\M=2$ and that $P_X^{(1)} \neq P_X^{(2)}$. In this case, for each $k\in\{1,2\}$, 
detector $k$'s exponent $\theta_k$  coincides with the Shimokawa-Han-Amari exponent \cite{shimokawa} achieved in a single-detector system with only  detector $k$. 
\end{corollary}

The scheme leading to Theorem~\ref{theorem-lower-bounds-power-exponents-general-hypotheses-positive-rates} applies binning as used in the Shimokawa-Han-Amari scheme \cite{shimokawa}. A better performance could be achieved in general if the scheme was replaced by the Heegard-Berger type-scheme in \cite{roy}. 

 We  now consider the case of zero rate, i.e., where the encoder's message takes value in $[1:\W]$, for some fixed finite $\mathsf{W}>0$. Assume for the rest of this section that for each $k\in[1:\K]$, each pair of pmfs $P_{XY_k}^{(m)}$ and $P_{XY_k}^{(m')}$ differs in at least one of the two marginals $P_X$ or $P_{Y_k}$. Further, 
assume  
\begin{equation}\label{eq:cond}
P_{XY_k}^{(k)}>0, \quad \forall k\in[1:\K], \qquad (x,y_k) \in \mc X \times \mc Y_k.
\end{equation}

Let $\L$ be the number of different pmfs in $\{P_X^{(1)}, \ldots, P_{X}^{(\M)}\}$. So, $1 \leq \L \leq \M$, where $\L=1$ means that pmfs $P_X^{(1)}, \ldots, P_{X}^{(\M)}$ all coincide. Let $P_{X,1}, \ldots, P_{X,\L}$ be these $\L$ distinct pmfs. The error exponent region depends on $\L$ and on $\W$. If $\W> \L$,  the exponent region is a $\K$-dimensional cube and in each dimension the exponent can be as large as in a system where only the corresponding detector is present.

\begin{proposition}[Exponents Region when $\W> \L$] 
	For any $\boldsymbol{\epsilon} \in (0,1)^{\K}$ and $\W > \L$, the exponents region $\mc E(0_{\mathsf{W}},\boldsymbol{\epsilon})$  coincides with the set of all nonnegative exponent vectors $\boldsymbol{\theta}$ satisfying:
\begin{equation}\label{eq:result-for-adequate-codebooks}
\theta_k \leq\min_{\substack{ m \in [1:\M]\\  m\neq k}  } \min_{\begin{array}{c}\pi_{XY_k}:\\ \pi_{X} = P_X^{(m)} \\ \pi_{Y_k} = P_{Y_k}^{(m)} \end{array}} D\Big(\pi_{XY_k}\|P_{XY_k}^{(k)}\Big)
\end{equation}
\end{proposition}
\begin{IEEEproof} The converse holds by  \cite[Theorem 5]{shalaby1992multiterminal} and because any detector cannot have a larger type-II error exponent as in a setup where it is the only detector and the transmitter can send a message of size $\W$ to this single detector. Achievability follows by  analyzing the following scheme. (Details of the analysis are omitted.) Fix a small $\mu>0$.\\ \underline{\textit{Encoder:}}  If the encoder observes $X^n=x^n$  and  $x^n \in \mathcal{T}_{\mu}( P_{X,\ell})$ for some index $\ell \in \{1,\ldots, \L\}$, then it sends $W= \ell$. Otherwise, it sends $W=\L+1$.  (Notice that since pmfs $P_{X,1},\ldots, P_{X,\L}$ are all distinct, for sufficiently small $\mu$, every $x^n\in\mathcal{X}^n$ belongs to only one  typical set $\mathcal{T}_{\mu}( P_{X,\ell})$.) \\ 	\underline{\textit{Detector $k$:}} If the received message $W \leq  \L$ and  for some $m\in[1:\M]$  the following two conditions hold:
	\begin{IEEEeqnarray}{rCl}
		P_X^{(m)} &= &P_{X,W}\\
	Y_k^n &\in& 	 \mathcal{T}_{\mu}\big( P_{Y_k}^{(m)}\big),
		\end{IEEEeqnarray}
detector $k$ sets $\hat{\mathcal{H}}_k=m$. Otherwise, it sets $\hat{\mathcal{H}}_k =k$. 
		\end{IEEEproof}

For each vector  $\mathbf{r}:=(r_1,\ldots, r_{\K-1})\in\mathbb{R}^{\K-1}$  and mapping $b \colon [1:\M] \to [1:\W]$, define a partition  $\phi_{b,1}(\mathbf{r}), \ldots, \phi_{b,\W}(\mathbf{r})$ of $\mathcal{P}(\mathcal{X})$ so that $\phi_{b,i}(\mathbf{r})$ contains   $P_X^{(m)}$ if and only if  $b(m)=i$, and $\phi_{b,i}(\mathbf{r})$ contains any  other type $\tilde{P}_X$ only if 
 \begin{IEEEeqnarray}{rCl} \label{eq:Dcond}
  	i &=& \textnormal{arg}\! \max_{\hspace{-3mm} j \in [1:\W]} \; 
  	\min_{\substack{\kappa \in [1:\mathsf{K}] } }  \nonumber \\
& &  \quad \quad	\bigg[ \min_{\substack{m \in [1:\M] \\ b(m)=j \\ m \neq \kappa} }   \min_{\substack{\pi_{XY_k}\colon \\ \pi_{X}\!=\!\tilde{P}_X  \\ \pi_{Y_{\kappa}}\!\!=\! P_{Y_{\kappa}}^{(m)}}}
  	\!\! D\Big(\pi_{XY_{\kappa}}\|P_{XY_{\kappa}}^{(\kappa)}\Big) 
		+\sum_{l=\kappa}^{\K}{r_l}\bigg]. \qquad  \label{eq:i}
  \end{IEEEeqnarray}

\begin{figure}[H]
	\centering
		\includegraphics[width=7cm]{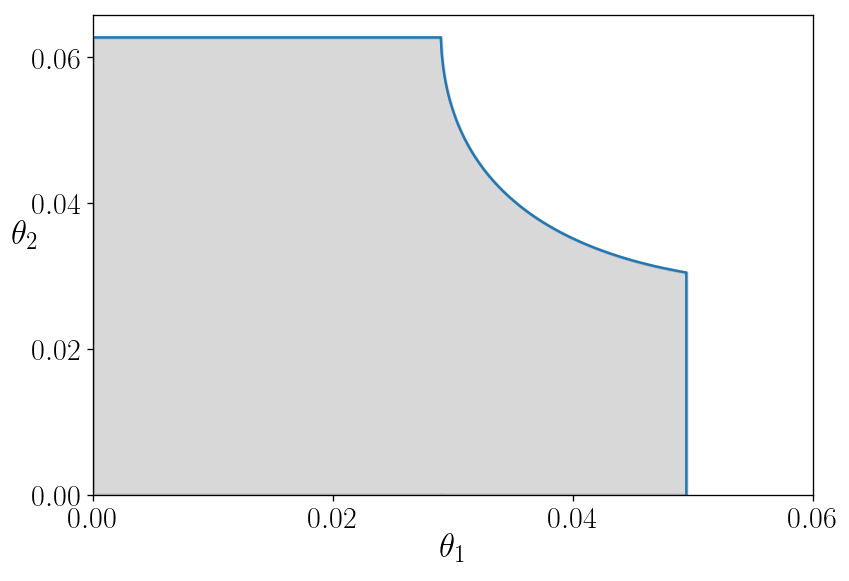}\vspace{-0.5cm}
	\caption{Exponent region of Example~\ref{ex1}, see \cite{simulation} for implementation details.}
	\label{fig-opposite-exponent-region-2}
\end{figure}

\begin{theorem}[Exponents Region when $\W\leq \L$]\label{theorem-zero-rate-result}
For any $\boldsymbol{\epsilon} \in (0,1)^\K$, the exponents region $\mc E(0_{\mathsf{W}},\boldsymbol{\epsilon})$   coincides with the set of  nonnegative exponent vectors $\boldsymbol{\theta}=(\theta_1,\ldots, \theta_\K)$ that   for some     $\mathbf{r}\in\mathbb{R}^{\K-1}$  and mapping $b \colon [1:\M] \to [1:\W]$ satisfy
\begin{equation}\label{optimality-result-for-bit-depletion-case}
\theta_k \leq  \min_{\substack{m \in [1: \M] \\
		m \neq k }} \min_{\substack{\pi_{XY_k}: \\[.3ex] \pi_{X} \in \phi_{b,b(m)}(\mathbf{r}) \\[.3ex] 
		\pi_{Y_k}=P_{Y_k}^{(m)} }}
\!\!\!{D\Big(\pi_{XY_k}\|P_{XY_k}^{(k)}\Big)}, \!\quad k\in[1:\K].
\end{equation}
\end{theorem}
\begin{IEEEproof}
See Section~\ref{secVI-subsecC}.
\end{IEEEproof}
 Given the mapping $b$, the theorem  describes an optimal choice of the partition $\phi_{b,1}(\mathbf{r}), \ldots, \phi_{b,\W}(\mathbf{r})$. It remains to optimize over the choice $b$. A similar phenomenon is encountered in the characterization of the optimal error exponent of the single-detector composite hypothesis testing problem  in    \cite{shalaby1992multiterminal} and in the characterization of the optimal set of type-I \emph{and} type-II error exponents in \cite{han_kobayashi}.
\begin{example} \label{ex1}Consider a setup where $\K=2$, $\M=3$ and $\W=2$ and where $X, Y_1, Y_2$ are binary with pmfs 
\begin{align}
\left\{\begin{array}{cc} P_{X Y_k}^{(1)}(0,0) = 0.30 & P_{X Y_k}^{(1)}(0,1) =0.23 \\ P_{X Y_k}^{(1)}(1,0) =0.27 & P_{X Y_k}^{(1)}(1,1) =0.20 \end{array} \right. \\
 \left\{\begin{array}{cc} P_{X Y_k}^{(2)}(0,0) =0.14 & P_{X Y_k}^{(2)}(0,1) =0.29 \\ P_{X Y_k}^{(2)}(1,0) =0.31 & P_{X Y_k}^{(2)}(0,1) =0.26 \end{array} \right. \\
\left\{ \begin{array}{cc} P_{X Y_k}^{(3)}(0,0) =0.52 & P_{X Y_k}^{(3)}(0,1) =0.18 \\ P_{X Y_k}^{(3)}(1,0) =0.23 & P_{X Y_k}^{(3)}(1,1) =0.07. \end{array}\right. 
 \end{align}
\end{example}
The exponent region corresponding to this example is depicted in Fig.~\ref{fig-opposite-exponent-region-2}. It is non-convex. (Notice that time-sharing arguments cannot be applied to convexify the region.)
\vspace{-0.4cm}

\section{Composite Hypothesis Testing}

Consider now the related \emph{composite testing} setup in which each detector aims at maximizing the error exponent that is associated to a \emph{set} of hypothesis.  For each $k \in [1:\K]$, let $\mathcal{S}_k$ be a subset of $[1:\M]$. Detector $k$ declares $\hat{\mc H}_k=0$ to indicate that $\mc H \in \mc S_k$, and it declares $\hat{ \mc H} =m$ to indicate that $\mc H=m$ for any hypothesis $m\in [1:\M]\backslash \mc S_k$. Specifically, it produces $\hat{\mc H}_k$ as in  \eqref{eq:decide} but using  a decision function of the form
\begin{equation}
g_{k,n} \colon [1:\W] \times \mathcal{Y}^n \to \{0\} \cup ( [1:\M] \backslash \mc S_k).
\end{equation}
Type-I error probabilities are defined as in the previous section. The type-II error probability at detector $k$ is defined as 
\begin{IEEEeqnarray}{rCl}
\gamma_{k,n}:= \min_{ \xi \in \mathcal{S}_k}  \text{Pr}\big\{\hat{\mc H}_k \neq 0 \big|\mc H =\xi \}.
\end{IEEEeqnarray}

The  exponents region is defined  as in Definitions~\ref{def:1} and ~\ref{def:region}, except that \eqref{eq-definition-constant-constraints-typeI-errors} and \eqref{eq:thetak} needs to be replaced by:

\begin{equation}
 \varlimsup_{n\to \infty} \alpha_{k,m,n} \leq \epsilon_k, \quad \forall m\in[1:\M]\backslash {\mc S}_k,\end{equation}
\begin{equation}
\eta_k \leq \varliminf_{n\to \infty} -\frac{1}{n} \log \gamma_{k,n},
\end{equation}
and  symbols $\theta$ and $\boldsymbol{\theta}$ have to be replaced by $\eta$ and $\boldsymbol{\eta}:=(\eta_1,\ldots, \eta_K)$. 

In this section, we allow for any number of hypotheses $\M \geq 2$, even  $\M < \K$. 
The result from the previous section, can be extended to  this composite hypothesis testing scenario. More specifically, the achievability parts of the following Theorem~\ref{thm3}  and 
Proposition~\ref{prop2} rely on the coding and testing schemes described previously, but where each detector simply declares $\hat{\mc H}_k =0$ instead of $\hat{\mc H}_k= m$ when $m\in \mc S_k$. The achievability proof of Theorem~\ref{thm4} necessitates further changes.
\begin{theorem}[Achievability under Positive Rate]\label{thm3}
	Given  $R > 0$ and $\boldsymbol{\epsilon}$ in $(0,1)^\K$, region $\mathcal{E}(R, \boldsymbol{\epsilon})$ contains all nonnegative vectors $\boldsymbol{\theta}=( \theta_1, \theta_2,\ldots,\theta_\K)$  that  satisfy the following two conditions for a tuple $\big(P_{U|X}^{(1)}, \ldots, P_{U|X}^{(\M)}\big)\in \mathcal{U}(R)$:
	\begin{align}\label{eq:eta1}
	& \eta_k \leq \min_{\substack{(\xi, m) \colon \\ \xi \in \mc S_k \\ m  \in [1:\M] \backslash \mc S_k} }\;\;
	\min_{\substack{ \pi_{UXY_k} \colon  \\  \pi_{UX}  = P^{(m)}_{UX}  \\ \pi_{UY_k} = P_{UY_k}^{(m)}  }} 
	D\left(\pi_{UXY_k}||P^{(m)}_{U|X}P_{XY_k}^{(\xi)}\right)
	\end{align}
	and 
	\begin{align}\label{eq:eta2}
 \eta_k& \leq \min_{\substack{(\xi, m) \colon \\ \xi \in \mc S_k \\ m  \in [1:\M] \backslash \mc S_k} }\;\; \min_{\substack{ \pi_{UXY_k} \colon  \\  \pi_{UX}  = P^{(m)}_{UX}  \\ \pi_{Y_k} = P_{Y_k}^{(m)} \\H_{P^{(m)}}(U|Y_k)\leq H_{\pi}(U|Y_k)}} \nonumber\\[1.2ex] &\Big[ D\left(\pi_{UXY_k}||P_{U|X}^{(m)}P_{XY_k}^{(\xi)}\right)  +R- I_{P^{(m)}}( U; X|Y_k) \Big].
	\end{align}
\end{theorem}
\begin{IEEEproof}
	Omitted.
	\end{IEEEproof}
This result differs from its simple-hypothesis testing counterpart in Theorem~\ref{theorem-lower-bounds-power-exponents-general-hypotheses-positive-rates} only in the first minimization of \eqref{eq:eta1} and \eqref{eq:eta2}. A similar statement applies to the following result.

Consider zero-rate communication. Assume \eqref{eq:cond} and that  any two $P_{XY_k}^{(m)}$ and $P_{XY_k}^{(m')}$ differ in at least one of the  marginals.

\begin{proposition}[Exponents Region when $\W> \L$]\label{prop2}
	For any $\boldsymbol{\epsilon} \in (0,1)^{\K}$ and $\W > \L$, the exponents region $\mc E(0_{\mathsf{W}},\boldsymbol{\epsilon})$  coincides with the set of all nonnegative exponent vectors $\boldsymbol{\theta}$ satisfying:
	\begin{equation}
\eta_k \leq \min_{\substack{(\xi, m) \colon \\ \xi \in \mc S_k \\ m  \in [1:\M] \backslash \mc S_k} }\;\; \min_{\begin{array}{c}\pi_{XY_k}:\\ \pi_{X} = P_X^{(m)} \\[1ex] \pi_{Y_k} = P_{Y_k}^{(m)} \end{array}} D\Big(\pi_{XY_k}\|P_{XY_k}^{(\xi)}\Big)
	\end{equation}
\end{proposition}
\begin{IEEEproof}
Omitted.
\end{IEEEproof}


\begin{theorem}[Exponents Region when $\W\leq \L$]\label{thm4}
	For any $\boldsymbol{\epsilon} \in (0,1)^\K$ the exponents region $\mc E(0_{\mathsf{W}},\boldsymbol{\epsilon})$   coincides with the set of  nonnegative exponent vectors $\boldsymbol{\theta}=(\theta_1,\ldots, \theta_\K)$ that for a 
	partition $\psi_1,\ldots, \psi_\W$ of $\mathcal{P}(\mc X)$  and a mapping $b \colon [1:\M] \to [1:\W]$ satisfy 
	\begin{equation*}
\eta_k \leq \min_{\substack{(\xi, m) \colon \\ \xi \in \mc S_k \\ m  \in [1:\M] \backslash \mc S_k} }\;\; \min_{\substack{\pi_{XY_k}: \\[.3ex] \pi_{X} \in \psi_{b(m)} \\[.3ex] 
			\pi_{Y_k}=P_{Y_k}^{(m)} }}
	\!\!\!{D\Big(\pi_{XY_k}\|P_{XY_k}^{(\xi)}\Big)}, \!\quad k\in[1:\K].
	\end{equation*}
\end{theorem}
\begin{IEEEproof}
Omitted.\end{IEEEproof}
The optimal partition $\psi_1,\ldots, \psi_\W$ of $\mc P(\mc X)$ in above Theorem~\ref{thm4} is in general different from the partitions $\phi_{b,1}(\mathbf{r}), \ldots,$ $ \phi_{b,\W}(\mathbf{r})$ introduced for the simple hypothesis testing setup.  This shows suboptimality of the naive strategy of applying the coding and testing scheme for the simple hypothesis testing problem and then declaring $\hat{\mc H}_k=0$ at a detector $k$ if the outcome of the test lies in $\mathcal{S}_k$ and declaring the produced hypothesis otherwise.
\section{Proof Of Theorem \ref{theorem-lower-bounds-power-exponents-general-hypotheses-positive-rates}}~\label{secV}

Fix $\mu>0$, a sufficiently large blocklength $n$ and  conditional pmfs $P_{U|X}^{(1)}, \ldots, P_{U|X}^{(\M)}$ so that \eqref{eq:rate} and \eqref{eq:same} are satisfied.   Define the joint pmfs $P^{(m)}_{UXY}=P^{(m)}_{XY}P^{(m)}_{U|X}$ 
and the nonnegative rate $R'$ such that for each $m\in[1:\M]$:
\begin{align}
R+R'&= I_{P^{(m)}}(X;U)+\mu,\label{cas1binp2pb}\\
R' &< I_{P^{(m)}}(Y_k;U), \quad k\in[1:\K]\backslash\{m\}.\label{rateKLH} 
\end{align}

\underline{\textit{Code Construction}:} For each $m\in[1:\M]$, construct a random codebook
\begin{align}
\mathcal{C}_{U,m} & =\{u_m^n(\tilde{w}, \ell)\colon \tilde{w} \in\{1,...,\lfloor e^{nR}\rfloor \}, \ell\in\{1,...,\lfloor e^{nR'}\rfloor \}\},\nonumber
\end{align}
by drawing all entries  i.i.d. according to $P_U^{(m)}$.
\vspace{2mm}

\underline{\textit{Sensor}:} Given that it observes the sequence $X^n=x^n$, the sensor looks for indices $(m,\tilde{w},\ell)$ such that 
\begin{align}\label{eq:typical_encoding}
(u_m^n(\tilde{w},\ell),x^n)\in \mathcal{T}_{\mu/2}(P^{(m)}_{UX}).
\end{align} 
If successful, it picks one of these indices uniformly at random and sends the pair $(m,\tilde{w})$  over the noise-free bit pipe. Otherwise, it sends $W=(0,1)$. (Notice that sending index $m$ takes only $\lceil \log \M \rceil$ bits and is thus of zero rate.)
\vspace{2mm}

\underline{\textit{Detector $k$}:} Assume that detector $k$  observes $Y_k^n=y_k^n$. If it obtains message $W=(0,1)$, then it declares   $\hat{\mathcal{H}}_k=k$. Otherwise, it parses message $W$ into the pair $(m,\tilde{w})$, and picks an index  $\ell'\in \{1,...,\lfloor e^{nR'}\rfloor \}$ such that for  all $\tilde{\ell} \in\{1,\ldots, \lfloor e^{nR'}\rfloor \}$:
\begin{align}
H_{\text{tp}(u_m^n(\tilde{w}, \ell'),y_k^n)}(U|Y_k)\leq  H_{\text{tp}(u_m^n(\tilde{w}, \tilde{\ell} ),y_k^n)}(U|Y_k).
\end{align}
It then checks whether 
\begin{equation}(u_m^{n}(\tilde{w},\ell'),y_k^n) \in \mathcal{T}_{\mu}(P_{UY}^{(m)}).
\end{equation}
If this test is successful, detector $k$ declares $\hat{\mathcal{H}}_k=m$. Otherwise, $\hat{\mathcal{H}}_k=k$.

\underline{\emph{Error analysis:}} By the same technical arguments as in the single decision center setup, see \cite[Proof of Thm. 4, App. H]{sadaf_cascade}. 

\section{Proof of Theorem \ref{theorem-zero-rate-result} }~\label{secVI-subsecC}
{\emph{Achievability:}} Fix  $\mu>0$ and  a function $b \colon [1:\M] \to [1:\W]$.  Fix also a vector  $\mathbf{r}=(r_1,\ldots, r_{\K-1})\in \mathbb{R}^{\K-1}$ and define the sets  $\phi_{b,1,\mu}(\mathbf{r}), \ldots, \phi_{b,\W,\mu}(\mathbf{r})$ so that they partition $\mathcal{P}^n(\mathcal{X})$, so that the types of the sequences in $\mathcal{T}_{\mu}(P_X^{(m)})$ belong to subset $ \phi_{b, b(m),\mu}(\mathbf{r})$, and so that   any other type $\tilde{P}_X$ is in $\phi_{b, b(m),\mu}(\mathbf{r})$ only if \eqref{eq:Dcond} holds. 

\underline{\textit{Sensor}:} Given that it observes the sequence $X^n=x^n$, the sensor sends $W=i$ if  $\textnormal{tp}( X^n) \in \phi_{b,i,\mu}(\textbf{r})$. 

\underline{\textit{Detector $k$:}} Given that it observes $Y_k^n=y_k^n$ and receives message $W=w$ from the sensor, detector $k$ checks whether for some $m\in[1:\M]$ the following two conditions hold: 
\begin{IEEEeqnarray}{rCl}
	b(m) = w  \qquad \textnormal{and} \qquad 
y_k^n \in \mathcal{T}_\mu\big( P_{Y_k}^{(m)}\big). 
	 \end{IEEEeqnarray} 
	 If such an index $m$ exists, Detector $k$ declares $\hat{\mathcal{H}}_k=m$. If none or multiple exist, it declares 
$\hat{\mathcal{H}}_k=k$.

\underline{\textit{Error analysis:}} Notice first that because all pmfs $P_{XY_k}^{(1)}, \ldots,$ $P_{XY_k}^{(\M)}$ differ in at least one of the two marginals, for sufficiently small  $\mu$,  when $X^n \in\mc T_{\mu}(P_X^{(m)})$ and $Y_k^n \in\mc T_{\mu}(P_{Y_k}^{(m)})$, then Detector~$k$ declares $\hat{\mc H}_k =m$. Thus, by the weak law of large numbers, for each $k\in [1:\K]$ and sufficiently large $n$:
\begin{IEEEeqnarray*}{rCl}
\alpha_{k,m,n} & \leq & 1- \Pr \Big[ (X^n,Y^n) \in \mc T_{\mu}(P_X^{(m)}) \times \mc T_{\mu}(P_{Y_k}^{(m)})\Big] \leq   \epsilon_k.
\end{IEEEeqnarray*}
Define now for each $k\in [1:\K]$, $m\in[1:\M]$, and $\mathbf{r}\in\mathbb{R}^{\K-1}$: 
\begin{IEEEeqnarray}{rCl}
\mc A_{k,m,\mu}(\mathbf{r})&: =&\Big \{ (x^n,y_k^n)\colon \quad   \tp{x^n} \in \phi_{b,b(m),\mu}(\textbf{r}), \nonumber \\
 & & \hspace{2.4cm}\textnormal{and} \qquad  y_k^n \in \mc T_{\mu}(P_{Y_k}^{(m)})\Big\}.\IEEEeqnarraynumspace
\end{IEEEeqnarray}
The type-II error probability at detector $k$ satisfies: 
\begin{IEEEeqnarray}{rCl}
\beta_{k,n} & \leq &\Pr \Big[ (X^n,Y_k^n)\in  \bigcup_{ \substack{ m \in [1: \mathsf{M}] \\ m \neq k } }\mc A_{k,m,\mu}  \Big| \mc H=k\Big] \nonumber \\
& \leq  & \min_{\substack{ m \in [1: \mathsf{M}] \\ m \neq k } } \; \min_{\substack{\pi_{XY_k} \colon \\ \pi_X \in  \phi_{b,b(m),\mu}(\textbf{r}) \\ \big|\pi_{Y_k} - P_{Y_k}^{(m)}\big|\leq \mu}} e^{-n \big(D\big( \pi_{XY_k} || P_{XY_k}^{(k)}\big)-\mu\big)}, \IEEEeqnarraynumspace
\end{IEEEeqnarray}
where the last inequality holds for sufficiently large values of $n$ and  by Sanov's theorem. 
Taking $\mu \to 0$ and $n\to \infty$ establishes the desired achievability result.

\emph{Converse:} Fix an integer number $n$ and define $\epsilon_{\min}$ to be the minimum component of $\boldsymbol{\epsilon}$. Since the sensor sends an index in $[1:\W]$,  for each $m \in [1:\M]$ there exists a partition $\mathcal{C}_1,\ldots, \mc C_\W$ of $\mathcal{X}^n$ and subsets $\mc F_1^{k,m}, \ldots,  \mc F_\W^{k,m}\subseteq \mathcal{Y}_k^n$ so that:
\begin{equation}
\hat{\mathcal{H}}_k =m \quad \Longleftrightarrow \quad  (X^n, Y_k^n) \in \left(\bigcup_{i=1}^\mathsf{W}{\mc C_i \times \mc F_i^{k,m}}\right).
\end{equation}
For each $i \in [1:\W]$, define the set 
\begin{IEEEeqnarray}{rCl}\label{eq:phin}
\phi_{i,n} := \bigg\{  \tilde{P}_X \in \mathcal{P}(\mathcal{X}) \colon \;  \tilde{P}_X^{\otimes n} \big[ X^n \in \mc C_i  \big] \geq \frac{1-\epsilon_{\min}}{\W}  \bigg\}.\IEEEeqnarraynumspace
\end{IEEEeqnarray}
Since the sets $\mathcal{C}_1,\ldots, \mc C_\W$ form a partition of $\mathcal{X}^n$ and since for each $\tilde{P}_X\in\mathcal{P}(\mc X)$ it holds that $ \tilde{P}_X^{\otimes n} \big[ X^n \in \mc X^n \big] =1$, the  subsets $\phi_{1,n}, \ldots, \phi_{\W,n}$ cover the set $\mathcal{P}(\mathcal{X})$. 
For a similar reason and  by the constraint on the type-I error probabilities, for each $k\in[1:\K]$ and $m\in [1:\M]$ with $m\neq k$, there exists an index $i(m,k)\in[1:\W]$ so that:
\begin{subequations}
	\label{type-I-cobstraint}
\begin{IEEEeqnarray}{rCl}
 {P_X^{(m)\otimes n}}\big[ X^n\in \mc C_{i(m,k)} \big] &\geq& \frac{1- \epsilon_{\min}}{\mathsf{W}}, \\
  P^{(m)\otimes n}_{Y_k}\big[ Y_k^n \in \mc F_{i(m,k)}^{k,m}\big] &\geq& \frac{1- \epsilon_{\min}}{\mathsf{W}}.
\end{IEEEeqnarray}
\end{subequations}
Notice  that  \eqref{type-I-cobstraint} implies that $P^{(m)}_X \in \phi_{i(m,k)}$. 
Combining \eqref{type-I-cobstraint} with the definition of $\phi_{i(m,k),n}$ in \eqref{eq:phin} and  with  \cite[Theorem 3]{shalaby1992multiterminal} yields that  for any $\mu>0$ and sufficiently large $n$:
\begin{align*}
\text{Pr}[ \hat{\mc H}_k=m | \mc H =k  ]\geq  \max_{ \substack{\pi_{XY_k}: \\ \pi_{X} \in \phi_{i(m,k),n} ,\\  \pi_{Y_k}=P_{Y_k}^{(m)}}} e^{-n \big(D\big( \pi_{XY_k} \| P^{(k)}_{XY_k}\big) + \mu\big)}. 
\end{align*}
Thus, 
\begin{IEEEeqnarray}{rCl}
\lefteqn{\text{Pr}[ \hat{\mc H}_k \neq k | \mc H =k  ] \nonumber} \quad \\
 & \geq \textnormal{exp} \Big( - n  \min_{\substack{m\in[1:\M] \\ m\neq k}} \min_{ \substack{\pi_{XY_k}: \\ \pi_{X} \in \phi_{i(m,k),n} ,\\  \pi_{Y_k}=P_{Y_k}^{(m)}}}  D\Big( \pi_{XY_k} \| P^{(k)}_{XY_k} \Big) - n \mu  \Big) \nonumber 
\vspace{-0.2cm}
\end{IEEEeqnarray}
\noindent Taking $\mu \to 0$ and $n \to \infty$, by continuity  we can conclude that if the exponent vector $\boldsymbol{\theta}=(\theta_1,\ldots, \theta_\K)$ is achievable, then there exist subsets $\phi_1,\ldots, \phi_\W$ so that 
\begin{IEEEeqnarray}{rCl}\label{eq:boundu}
\theta_{k} \leq \min_{\substack{m\in[1:\M]\\ m\neq k}} \min_{ \substack{\pi_{XY_k}: \\ \pi_{X} \in \phi_{b(m)} ,\\  \pi_{Y_k}=P_{Y_k}^{(m)}}}  D\big( \pi_{XY_k} \| P^{(k)}_{XY_k}\big) . \IEEEeqnarraynumspace
\vspace{-0.3cm}
\end{IEEEeqnarray}
where $b(m)$ is the index of the set $\phi_i$ containing $P_X^{(m)}$.
Since the upper bounds in \eqref{eq:boundu} become  looser when elements are removed from the sets $\phi_{i}$, the converse statement  remains valid by imposing that  $\phi_1,\ldots, \phi_{\W}$ form a partition of $\mathcal{P}(\mathcal{X})$.

Fix now $\mathbf{r}=(r_1,\ldots, r_{\K-1})\in \mathbb{R}^{\K-1}$ and consider only the  exponent vectors $\boldsymbol{\theta}=(\theta_1,\ldots, \theta_\K)$ that satisfy 
\begin{align} \label{constrained-space}
\theta_k &= \theta_{k+1} + r_k, \quad k\in\{1,\ldots,\K -1\}. 
\end{align}
Then  constraints  \eqref{eq:boundu} are loosest if  $\phi_{i}$ contains pmf $\tilde{P}_{X} \in \mathcal{P}(\mathcal{X})\backslash\{P_X^{(1)}, \ldots, P_X^{(\M)}\}$ only if condition \eqref{eq:i} is satisfied.  
\section*{Acknowledgement} 
M. Wigger was supported by
the ERC Grant CTO Com.
\bibliographystyle{IEEEtran}
\bibliography{IEEEabrv,./ISIT-2018-DraftV17-Pierre}

\end{document}